\documentclass[12pt,draftclsnofoot, onecolumn]{IEEEtran}
\usepackage{graphicx,cite,epsfig,amssymb,amsmath}
\usepackage{color,xcolor}
\usepackage{stmaryrd}
\usepackage{bbding}
\usepackage{subfigure}
\usepackage{setspace}
\usepackage{color}

\begin{document}
\begin{spacing}{1.75}


\title{Resource Allocation for Cost Minimization in Limited Feedback MU-MIMO Systems with Delay Guarantee}

\author{\normalsize
Xiaoming~Chen, Zhaoyang~Zhang, and Chau~Yuen
\thanks{This work was supported by the National Natural Science Foundation
of China (No. 61301102), Natural Science Foundation of Jiangsu
Province (No. BK20130820), the open research fund of National Mobile
Communications Research Laboratory, Southeast University (No.
2012D16) and the Doctoral Fund of Ministry of Education of China
(No. 20123218120022).}
\thanks{Xiaoming~Chen (e-mail: {\tt chenxiaoming@nuaa.edu.cn}) is
with the College of Electronic and Information Engineering, Nanjing
University of Aeronautics and Astronautics, and also with the
National Mobile Communications Research Laboratory, Southeast
University, China. Zhaoyang~Zhang (e-mail: {\tt
ning\_ming@zju.edu.cn}) is with the Department of Information
Science and Electronic Engineering, Zhejiang University, China.
Chau~Yuen (e-mail: {\tt yuenchau@sutd.edu.sg}) is with Singapore
University of Technology and Design, Singapore.}}

\date{\today}
\renewcommand{\baselinestretch}{1.5}
\thispagestyle{empty} \maketitle

\begin{abstract}
In this paper, we design a resource allocation framework for the
delay-sensitive Multi-User MIMO (MU-MIMO) broadcast system with
limited feedback. Considering the scarcity and interrelation of the
transmit power and feedback bandwidth, it is imperative to optimize
the two resources in a joint and efficient manner while meeting the
delay-QoS requirement. Based on the effective bandwidth theory, we
first obtain a closed-form expression of average violation
probability with respect to a given delay requirement as a function
of transmit power and codebook size of feedback channel. By
minimizing the total resource cost, we derive an optimal joint
resource allocation scheme, which can flexibly adjust the transmit
power and feedback bandwidth according to the characteristics of the
system. Moreover, through asymptotic analysis, some simple resource
allocation schemes are presented. Finally, the theoretical claims
are validated by numerical results.
\end{abstract}

\begin{keywords}
\begin{center}
MU-MIMO, delay-sensitive, resource allocation, cost minimization,
limited feedback.\end{center}
\end{keywords}
\IEEEpeerreviewmaketitle

\newpage

\section{Introduction}

Despite great progress in wireless communications during the past
several years, it is still a challenging task to support
delay-sensitive wireless services over time-varying fading channels
given the scarcity of  resources, such as wireless spectrum and
energy efficiency. Under these circumstances, efficient resource
allocation with delay guarantee (due to its wide applications in
video broadcasting or tele-conferencing) becomes a hot topic in
wireless research community \cite{DC1}~-\cite{DC3}. In addition,
advanced multiple antennas techniques have became the basic
assumption in many international communications standards, e.g. IEEE
802.11ac and LTE-A, in order to support next generation multimedia
communications. In particular, Multi-User MIMO (MU-MIMO) is a key
technology that has been adopted by IEEE 802.11ac and LTE-A due to
its capability to exploit the extra antenna at the base-station to
serve multiple users concurrently.

Resource allocation in MU-MIMO systems receives considerable
attentions due to its potential of improving spectral efficiency
remarkably by exploiting the unique spatial degree of freedom, as
seen in \cite{MU-MIMO1}~-\cite{MU-MIMO3} and references therein. As
suggested by previous study \cite{CSIamount}, the performance of
MU-MIMO downlink is closely related to the amount of channel state
information (CSI) at the base station (BS). For example, if the BS
has no CSI, it can only work with a fixed transmission scheme, which
is equivalent to the traditional point-to-point MIMO system
\cite{NoCSI}. If full CSI is available at the BS, dirty paper coding
can be used to approach the capacity of MU-MIMO downlink
\cite{FullCSI}. Between the two extreme cases, if the BS has partial
CSI, some effective preprocess techniques, such as zero-forcing
beamforming (ZFBF) \cite{ZFBF1} \cite{ZFBF2}, can be used to
partially cancel the inter-user interference. Thus, the amount of
CSI greatly impacts the performance of the MU-MIMO system.

Furthermore, the performance of the MU-MIMO system is also affected
by other resources, such as transmit power \cite{Chen12}, since both
the desired signal quality and the inter-user interference are the
functions of the transmit power. Note that in the MU-MIMO system,
power and feedback resources are interrelated for a given QoS
constraint, e.g. delay requirement. For a MU-MIMO system with
delay-QoS guarantee, the critical issue of the design of such joint
power and feedback resource allocation is to reveal the relationship
between the delay requirement and the involved resources. Based on
the availability of CSI at the BS, a variety of models have been
built to characterize the relationship between the delay requirement
and the involved resources, and several adaptive resource allocation
schemes have been derived \cite{DCSurvey}, such as equivalent rate
constraint \cite{ERCModel1} \cite{ERCModel2}, Lyapunov drift
\cite{LyapnovModel1} \cite{LyapnovModel2} and Markov decision
process (MDP) \cite{MDPModel1} \cite{MDPModel2} schemes.

All the aforementioned delay driven resource allocation schemes
consider average delay as the QoS requirement. In fact, for some
delay sensitive services, e.g. video and audio services, maximum
delay is of concern. Based on the large deviation principle,
effective bandwidth theory can be used to establish the relationship
between maximum delay and minimum serve rate for a given violation
probability \cite{EffectiveBandwidth1} \cite{EffectiveBandwidth2}.
Hence \cite{EffectiveBandwidth3} studied the power allocation with
maximum delay constraint in wireless systems according to the
effective bandwidth theory. Nearly all previous works on resource
allocation based on the effective bandwidth theory assume full CSI
at the BS. However, as mentioned earlier, the BS only has partial
CSI by consuming the feedback resource. To the best of our
knowledge, feedback resource allocation in the delay-sensitive
MU-MIMO downlink has not been well addressed.

In this paper, we focus on joint power and feedback resource
allocation with maximum delay guarantee in the MU-MIMO downlink
employing with ZFBF. Different from previous works, we consider
limited CSI feedback with quantization codebooks. The main
contribution of this paper lies in that we reveal the relation
between the violation probability with respect to a maximum delay
requirement, transmit power and feedback bandwidth based on the
effective bandwidth theory, and then derive an optimal joint power
and feedback resource allocation scheme by minimizing the resource
cost function while satisfying delay constraint. The main
contributions of this paper can be summarized as follows:
\begin{enumerate}
\item We reveal the relation between the violation probability
with respect to a maximum delay requirement, transmit power and
feedback bandwidth based on the effective bandwidth theory.

\item We design a framework of joint power and feedback resource
allocation in the delay-sensitive downlink MU-MIMO with limited
feedback, and propose an optimal joint resource allocation schemes.

\item We formulate a resource cost function as the sum of power cost
and feedback cost. By adjusting the relative cost factor according
to the characteristics of the considered system, we can obtain the
corresponding resource allocation results.

\item Through asymptotic analysis, we obtain two simple resource allocation
schemes in the interference-limited and noise-limited scenarios
respectively.
\end{enumerate}

The rest of this paper is organized as follows. In Section II, the
considered system model is briefly introduced, and then the adopted
transmission protocol and effective bandwidth theory are discussed.
We propose a joint transmit power and feedback bandwidth allocation
scheme with delay-QoS provisioning in Section III. In Section IV, we
derive two simple resource allocation schemes via asymptotic
analysis. Simulation results are presented in Section V, and we
conclude the whole paper in Section VI.

\textit{Notation}: We use bold upper (lower) letters to denote
matrices (column vectors), $(\cdot)^H$ to denote conjugate
transpose, $(\cdot)^T$ to denote matrix transpose,
$\|\textbf{x}\|_2$ to denote the $l_2$ norm of vector $\textbf{x}$,
$|y|$ to denote the absolute value of $y$ and $G^{'}(x)$ to denote
the differential of function $G(x)$ with respect to $x$. The acronym
i.i.d. means ``independent and identically distributed", pdf means
``probability density function" and cdf means ``cumulative
distribution function".

\section{System Model}
\begin{figure}[h] \centering
\includegraphics [width=0.8\textwidth] {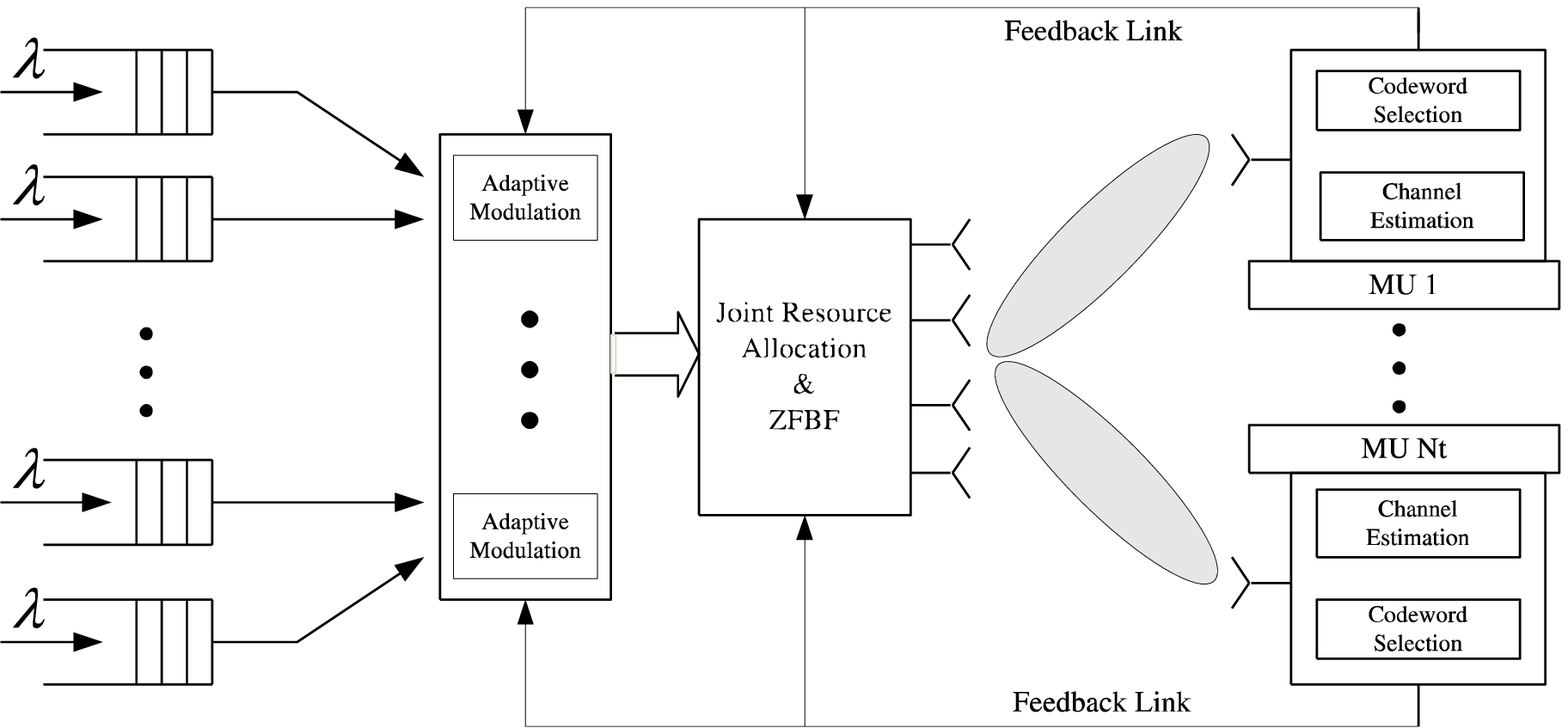}
\caption {The considered system block diagram.} \label{Fig1}
\end{figure}

Consider a homogeneous MU-MIMO downlink, which includes a base
station (BS) with $N_t$ antennas, and $N_t$ single antenna mobile
users (MUs), as shown in Fig.\ref{Fig1}. For tractability, it is
assumed that the downlink channels $\textbf{h}_k, k=1,\cdots,N_t$
from the BS to MUs are i.i.d. complex Gaussian random vectors with
zero mean and unit variance. All the channels are assumed to remain
unchanged during one time slot and fade independently slot by slot.
At the beginning of each time slot, MUs convey the corresponding CSI
to the BS based on quantization codebooks. All the codebooks are
designed in advance and stored at the BS and MUs. Assuming the
codebook of size $2^B$ at the $k$th MU is
$\mathcal{H}_k=\{\hat{\textbf{h}}_{k,1},\cdots,\hat{\textbf{h}}_{k,2^B}\},k=1,\cdots,N_t$,
the optimal quantization codeword selection criterion can be
expressed as
\begin{equation}
i=\arg\max\limits_{1\leq
j\leq2^B}|\hat{\textbf{h}}_{k,j}^{H}\tilde{\textbf{h}}_k|^2,\label{eqn1}
\end{equation}
where $\tilde{\textbf{h}}_k=\frac{\textbf{h}_k}{\|\textbf{h}_k\|}$
is the channel direction vector. Specifically, the optimal codeword
index $i$ is conveyed by the $k$th MU and $\hat{\textbf{h}}_{k,i}$
is recovered at the BS as the instantaneous CSI of the $k$th MU.

Based on the feedback information from the MUs, the BS designs the
optimal transmit beams $\textbf{w}_{k},k=1,\cdots,N_t$ by making use
of the ZFBF design method \cite{ZFBF1}. For the $k$th MU, the BS
first constructs its complementary channel matrix
\begin{equation}
{\hat{\textbf{H}}}_k=[\hat{\textbf{h}}_{1}, \cdots,
\hat{\textbf{h}}_{k-1}, \hat{\textbf{h}}_{k+1}, \cdots,
\hat{\textbf{h}}_{N_t}],\nonumber
\end{equation}
where $\hat{\textbf{h}}_{k-1}$ is the $(k-1)$th MU's optimal channel
quantization codeword. Taking singular value decomposition (SVD) to
${\hat{\textbf{H}}}_k$, if $\textbf{V}_{k}^{\perp}$ is the matrix
composed of the right singular vectors with respect to zero singular
values, then $\textbf{w}_k$ is a normalized vector spanned by the
space of $\textbf{V}_{k}^{\perp}$, so we have
\begin{equation}
\hat{\textbf{h}}_{u}^H\textbf{w}_k=0,\quad 1\leq k, u\leq N_t, u\neq
k.\nonumber
\end{equation}
It is assumed that $x_k$ is the desired normalized signal of the
$k$th MU, then its receive signal can be expressed as
\begin{equation}
y_k=\sqrt{\frac{P}{N_t}}\sum\limits_{u=1}^{N_t}\textbf{h}_k^H\textbf{w}_ux_u+n_k,\label{eqn2}
\end{equation}
where $P$ is the total transmit power of the BS, which is equally
allocated to the $N_t$ MUs. $n_k$ is the additive Gaussian white
noise with zero mean and variance $\sigma_n$ for all MUs. Hence, the
ratio of the received signal to interference and noise (SINR) for
the $k$th MU can be expressed as
\begin{eqnarray}
\rho_k&=&\frac{P/N_t|\textbf{h}_k^H\textbf{w}_k|^2}{\sigma_n^2+P/N_t
\sum\limits_{u=1,u\neq k}^{N_t}|\textbf{h}_k^H\textbf{w}_u|^2}\nonumber\\
&=&\frac{|\textbf{h}_k^H\textbf{w}_k|^2}{1/\gamma+\sum\limits_{u=1,u\neq
k}^{N_t}|\textbf{h}_k^H\textbf{w}_u|^2},\label{eqn3}
\end{eqnarray}
where $\gamma=\frac{P}{N_t\sigma_n^2}$ is the average transmit SNR
at the BS and $P/N_t\sum\limits_{u=1,u\neq
k}^{N_t}|\textbf{h}_k^H\textbf{w}_u|^2$ is the inter-user
interference. Although ZFBF is adopted, there are some residual
interference, since the BS only has the quantized CSI
$\hat{\textbf{h}}_k, k=1,\cdots, N_t$. The beam $\textbf{w}_u$ is
designed according to the criterion
$\hat{\textbf{h}}_k^H\textbf{w}_u=0$, so we have
$\textbf{h}_k^H\textbf{w}_u\neq0$, resulting in the residual
interference. Clearly, the more the feedback amount, the less the
residual interference. If the BS has full CSI, the interference can
be canceled completely due to
$\hat{\textbf{h}}_k=\tilde{\textbf{h}}_k$. It is worth pointing out
that the SINRs for all MUs have the similar expression in virtue of
the homogeneous characteristics.

\subsection{Transmission Protocol}
The data from high layer is organized in packet at the data link
layer. Each packet has a fixed number of bits $N_b$, including
packet header, payload and cycle redundant check (CRC). The arrived
data packets for each MU enter its unique buffer (with infinite
capacity) and wait for transmission at the physical layer. Within
each transmission duration $T_b$, the packets at the front of the
buffers are adaptively modulated respectively according to the
respective channel quality for the corresponding MU, namely SINR,
and then form $N_t$ independent data frames. Specifically, the whole
SINR range is partitioned into $N$ regions by $N+1$ SINR thresholds
$\Omega_n, n=0,\cdots,N$. For the $k$th MU, if the SINR of the
current frame $\rho_k$ satisfies the condition that
$\Omega_n\leq\rho_k<\Omega_{n+1}$, then the $n$th modulation format
is selected. The determination of SINR threshold depends on the
relationship between packet error rate (PER) and modulation mode.
Through theoretic analysis and numerical simulation, the expression
of PER in the form of modulation mode is given by \cite{AMC}
\begin{equation}
P_{e,n}(\rho)\approx\left\{\begin{array}{ll}
                1,& \mbox{ if $0<\rho<\gamma_{pn}$}\\
                a_n\exp(-g_n\rho), & \mbox{ if
                $\rho\geq\gamma_{pn}$} \end{array}
                \right. \label{eqn17}
\end{equation}
where $\gamma_{pn}$ is the cutoff SINR, below which the PER is
unacceptable. Notably, $a_n$, $g_n$ and $\gamma_{pn}$ are the
parameters dependent on the modulation mode $n$. For packet length
$N_b=1080$, these parameters for various modulation modes can be
found in Table \ref{Tab1}. It is assumed that the objective PER is
fixed as $P_{\textrm{obj}}$, a feasible method of determining the
$N+1$ SNR thresholds is given by
\begin{eqnarray}
\Omega_0&=&0,\nonumber\\
\Omega_n&=&\frac{1}{g_n}\ln\left(\frac{a_n}{P_{\textrm{obj}}}\right),\quad n=1,\cdots,N-1\nonumber\\
\Omega_{N}&=&\infty.\label{eqn18}
\end{eqnarray}
Due to the delay constraint of real-time services, the packet will
be discarded if its waiting time exceeds the upper bound on delay
$D_{\textrm{max}}$.

\subsection{Effective Bandwidth}
Since the pioneer work of Kelly \cite{EffectiveBandwidth2}, the
concept of effective bandwidth is widely used in wireless
communications together with queueing theory. Effective bandwidth is
defined as the characteristics of the source in bounds, limits and
approximations for various models of multiplexing under QoS
constraints. Specifically, in this paper, effective bandwidth is
defined as the minimum serve rate required by a stationary and
ergodic arrival process while fulfilling the constraint of maximum
waiting delay.

As mentioned earlier, the packet will be discarded if the waiting
time is greater than $D_{\textrm{max}}$. Thus, according to the
large deviation principle (LDP), the probability of dropping packet
caused by delay, so called violation probability, can be written as
\cite{EC}
\begin{eqnarray}
P_d(C)\approx\hat{\rho}\exp\left(-\delta(C)CD_{\textrm{max}}\right),\label{eqn7}
\end{eqnarray}
where $C$ is the serve rate, $\delta(C)=\max(s\geq0;\alpha(s)\leq
C)$ is the QoS exponent, the increasing function $\alpha(s)$ is the
so-called effective bandwidth of the traffic source and $s$ denotes
the space variable. $\hat{\rho}$ is the probability that the buffer
is nonempty. Let $S_n$ be the indicator of whether the $n$th packets
is in service ($S_n\in{0,1}$) and $M$ is the total number of packet,
then the approximate nonempty probability $\hat{\rho}$ can be
reckoned as
\begin{equation}
\hat{\rho}\approx\frac{1}{M}\sum\limits_{m=1}^{M}S_m.\label{eqn6}
\end{equation}
Given the maximum delay constraint $D_{\textrm{max}}$ and the
tolerable upper bound on violation probability
$\varepsilon=\hat{\rho}\exp\left(-\delta(C)CD_{\textrm{max}}\right)$,
we could determine the required minimum serve rate
$C_{\textrm{min}}$. For the Poisson arrival process, $\alpha(s)$ can
be computed as \cite{EffectiveBandwidth2}

\begin{equation}
\alpha(s)=\frac{\lambda}{s}\int_{0}^{\infty}\left(\exp(sx)-1\right)dF(x),\label{eqn8}
\end{equation}
where $F(x)$ is the cumulative distribution function (cdf) of the
packet size. Since all packets have the same size, $F(x)$ is a step
function at $x=N_b$. Thereby, we have
\begin{equation}
\alpha(s)=\frac{\lambda}{s}\left(\exp(sN_b)-1\right).\label{eqn9}
\end{equation}
By combining (\ref{eqn7}) and (\ref{eqn9}), we can obtain the
property of the required minimum served rate as follows.

\emph{Theorem 1}: When the average arrival rate $\lambda$ is large
enough, the required minimum served rate can be approximately
expressed as $C_{\textrm{min}}(\lambda,D_{\textrm{max}})\simeq
N_b\lambda-\frac{N_b(\ln\varepsilon-\ln\hat{\rho})}{2D_{\textrm{max}}}$.
In other words, $C_{\textrm{min}}(\lambda,D_{\textrm{max}})$
increases approximately linearly with $\lambda$.

\emph{Proof}: Please refer to Appendix I for the proof.

\emph{Corollary 1}: For high arrival rate, the gap $\Delta_C$
between $C_{\lambda,\textrm{min}}(D_{\textrm{max},1})$ and
$C_{\textrm{min}}(\lambda,D_{\textrm{max},2})$ is a constant
regardless of $\lambda$, where $D_{\textrm{max},1}$ and
$D_{\textrm{max},2}$ are two arbitrary maximum delay constraints.

Because $\alpha(s)$ is an increasing function of $s$,
$\delta(C)=\max(s\geq0;\alpha(s)\leq C)$ has a unique solution in
the form of $C$. In this context, we could derive the required
minimum serve rate $C_{\textrm{min}}(\lambda,D_{\textrm{max}})$ as a
function of the QoS requirement $D_{\textrm{max}}$ and $\varepsilon$
for an arbitrary arrival rate $\lambda$ by solving the equations
(\ref{eqn7}) and (\ref{eqn9}). Similarly, given $C$ and
$D_{\textrm{max}}$, we can also obtain the corresponding
$\varepsilon$ easily.

\section{Joint Resource Allocation with Delay Guarantee}
In this section, we focus on joint transmit power and feedback
bandwidth allocation while satisfying the delay constraint in a
MU-MIMO downlink. Note that in this paper transmit power is the same
for all MUs and only the total power is regulated, since equal power
allocation is optimal in the statistical sense and can avoid a large
amount of information feedback in a multiuser system. Considering
the scarcity of the two resources in practical systems, we expect to
minimize the total resource utilization. Hence, we set the
optimization objective as minimizing the following resource cost
function:
\begin{equation}
\eta=\varphi P+\psi N_tB,\label{eqn20}
\end{equation}
where $\varphi$ (cost per watt) and $\psi$ (cost per bit) are the
cost factors of transmit power $P$ and feedback amount $N_tB$,
respectively. The cost factors depends on the characteristics of the
considered system. For example, in a power-limited system, more
feedback amount should be used to decrease the consumption of
transmit power. Otherwise, if the system is feedback limited, it is
better to use more transmit power. By changing the relative cost
factor $\xi=\frac{\varphi}{\psi}$, we can character the different
systems. As a simple example, $\xi>>1$ denotes the power-limited
system, while $\xi<<1$ denotes the feedback-limited system. Since
the two resources are independent of each other, we model the total
cost as the linear sum of the two resource costs.

As analyzed earlier, given data arrival rate, the violate
probability with respect to a maximum delay constraint is a function
of the serve rate. Thus, based on adaptive modulation, the average
violation probability for the $k$th MU with transmit power $P$ and
codebook size $B$ can be computed as
\begin{eqnarray}
\bar{\varepsilon}_k(P,B)&=&\sum\limits_{n=0}^{N}P_d(C_n)P_r(C_n)\nonumber\\
&=&\sum\limits_{n=0}^{N}P_d(C_n)\bigg(F_{\rho_k}(\Omega_{n+1})-F_{\rho_k}(\Omega_n)\bigg),
\label{eqn14}
\end{eqnarray}
where $F_{\rho_k}(x)$ is the cumulative distribution function (cdf)
of SINR $\rho_k$, and
$P_r(C_n)=F_{\rho_k}(\Omega_{n+1})-F_{\rho_k}(\Omega_n)$ is the
probability that the $n$th serve rate or modulation mode is
selected. Assuming the downlink bandwidth is $W$, then we have
$C_n=nW/N_b$. For the $n$th modulation mode or given the serve rate
$C_n$, the nonempty probability $\rho(C_n)$ and the QoS exponent
$\delta(C_n)$ are fixed based on (\ref{eqn7}) and (\ref{eqn9}), so
the violation probability $P_d(C_n)$ is a constant. Following
\cite{ZFBF1} \cite{ZFBF2}, the cdf of $\rho_k$ based on limited
feedback ZFBF can be expressed as
\begin{equation}
F_{\rho_k}(x)=1-\frac{\exp\left(-x/\gamma\right)}{\left(1+\theta
x\right)^{N_t-1}},\label{eqn15}
\end{equation}
where $\theta=2^{-\frac{B}{N_t-1}}$. Substituting (\ref{eqn15}) into
(\ref{eqn14}), we have
\begin{eqnarray}
\bar{\varepsilon}_k(P,B)&=&\sum\limits_{n=0}^{N}P_d(C_n)\left(\frac{\exp\left(-\Omega_n/\gamma\right)}{\left(1+\theta
\Omega_n\right)^{N_t-1}}-\frac{\exp\left(-\Omega_{n+1}/\gamma\right)}{\left(1+\theta
\Omega_{n+1}\right)^{N_t-1}}\right)\nonumber\\
&=&P_d(C_0)-\sum\limits_{n=1}^{N}\bigg(P_d(C_{n-1})-P_d(C_{n})\bigg)\frac{\exp\left(-\Omega_n/\gamma\right)}{\left(1+\theta
\Omega_n\right)^{N_t-1}}.\label{eqn19}
\end{eqnarray}

Therefore, cost-minimizing joint resource allocation with delay
guarantee is equivalent to the following optimization problem
\begin{eqnarray}
J_1: \min\limits_{P,B} &&\eta=\varphi P+\psi N_tB\label{eqn21}\\
s.t.\quad &&\bar{\varepsilon}_k(P,B)\leq\varepsilon_0,
k=1,\cdots,N_t\nonumber\\
&&P\leq P_0\nonumber\\
&&B\leq B_0,\nonumber
\end{eqnarray}
where $\varepsilon_0$, $P_0$ and $B_0$ are the constraints on
violation probability, transmit power and feedback bandwidth,
respectively. Since $B$ is an integer variable, $J_1$ is a mixed
integer programming problem, it is difficult to obtain a closed-form
expression for the optimal $P$ and $B$. Intuitively, the optimal
algorithm is to compute the transmit power by letting
$\bar{\varepsilon}_k(P,B)=\varepsilon_0$ for a given $B$. Then,
search the optimal resource combination with the minimum resource
cost by scaling $B$ from 1 to $B_0$, namely the exhaustive search
algorithm. In fact, $J_1$ can be transformed as a general
optimization problem if $B$ is relaxed to a nonnegative real
variable, so that it can be solved by some optimization softwares,
such as the \emph{Lingo}. Assuming $P^{\dag}$ and $B^{\dag}$ are the
optimal solutions of the relaxed optimization problem, where
$B^{\dag}$ may not be an integer. Under this condition, we take
$B_c=\lceil B^{\dag}\rceil$ and $B_f=\lfloor B^{\dag}\rfloor$ as the
two candidates, where $\lceil B^{\dag}\rceil$ and $\lfloor
B^{\dag}\rfloor$ mean the smallest integer not less than $B^{\dag}$
and the largest integer not greater than $B^{\dag}$, respectively.
Given $B_c$, we could get the optimal transmit power $P_c$ by
letting $\bar{\varepsilon}_k(P_c,B_c)=\varepsilon_0$ and the
corresponding resource cost function $\eta(P_c, B_c)$. Similarly, we
also could obtain $P_f$ and $\eta(P_f, B_f)$ based on $B_f$. Then,
if $\eta(P_c, B_c)<\eta(P_f, B_f)$, we take $(P_c, B_c)$ as the
final resource combination. Otherwise, $(P_f, B_f)$ is selected.
Thus, we proposed a joint resource allocation scheme based on the
above idea. First, we derive the optimal feedback amount and
transmit power based on the relaxed optimization problem by using
the \emph{Lingo}. Then, we round the feedback amount to two nearest
integers, and compute the maximum transmit power while fulfilling
delay constraint. Finally, by comparing the corresponding resource
cost, the resource combination with the smallest cost is selected.
The joint resource allocation scheme can be described as follows
\rule{16cm}{1pt}
\begin{enumerate}
\item Initialization: given $N_t$, $N$, $P_0$, $B_0$, $\varphi$ and
$\psi$.

\item Relax $B$ to a nonnegative real number and derive $P^{\dag}$ and
$B^{\dag}$ by the \emph{Lingo}.

\item Let $B_c=\lceil B^{\dag}\rceil$ and $B_f=\lfloor B^{\dag}\rfloor$.
Compute $P_c$ satisfying
$\bar{\varepsilon}_k(P_c,B_c)=\varepsilon_0$ and $P_f$ satisfying
$\bar{\varepsilon}_k(P_f,B_f)=\varepsilon_0$.

\item Let $\eta(P_c, B_c)=\varphi P_c+\psi N_tB_c$ and $\eta(P_f, B_f)=\varphi P_f+\psi
N_tB_f$. If $\eta(P_c, B_c)<\eta(P_f, B_f)$, $(P_c, B_c)$ is the
final resource combination. Otherwise, $(P_f, B_f)$ is the required
one.

\end{enumerate}

\rule{16cm}{1pt}

Interestingly, it is found that although the proposed algorithm is
derived based on the relaxed optimization problem, it is also
optimal together with the exhaustive search algorithm. The complete
proof is given by Appendix II.

\section{Asymptotical Analysis}
In this section, we analyze the asymptotical characteristics of
violation probability in some special cases. Based on the insight
from the analysis, we can then derive some simple resource
allocation schemes that are optimal asymptotically.

\subsection{Interference-Limited Case}
If the variance of noise is quite small, e.g. in the noise-free
scenario, the noise can be negligible with respect to the inter-user
interference, namely the interference-limited case. Under such a
condition, the received SINR of the $k$th MU can be approximated as
\begin{equation}
\rho_k\approx\frac{|\textbf{h}_k^H\textbf{w}_k|^2}{\sum\limits_{u=1,u\neq
k}^{N_t}|\textbf{h}_k^H\textbf{w}_u|^2}.\label{eqn24}
\end{equation}
According to the analysis in previous section, we have the cdf of
$\rho_k$ in this case as
\begin{equation}
F_{\rho_k}(x)=1-(1+\delta x)^{-(N_t-1)}.\label{eqn25}
\end{equation}
Thereby, the joint resource allocation can be described as the
following optimization problem
\begin{eqnarray}
J_2: \min\limits_{P,B} &&\eta=\varphi P+\psi N_tB\label{eqn25}\\
s.t.\quad
&&P_d(C_0)-\sum\limits_{n=1}^{N-1}\bigg(P_d(C_{n-1})-P_d(C_{n})\bigg)\left(1+\Omega_n2^{-\frac{B}{N_t-1}}\right)^{-(N_t-1)}\leq\varepsilon_0\nonumber\\
&&P\leq P_0\nonumber\\
&&B\leq B_0.\nonumber
\end{eqnarray}

It is found that the first constraint is independent of $P$, this is
because the effects of $P$ on the desired signal and interference
are canceled out each other when the noise is ignored. From the
perspective of the optimization, it seems that $P=0$ is the optimal
solution to $J_2$. However, in practical systems, a minimum transmit
power is required to maintain the communications. In other words,
$P$ is set as the required minimum transmit power. For the optimal
codebook size, we first compute the $B^{\dag}$ satisfying
$P_d(C_0)-\sum\limits_{n=1}^{N-1}\bigg(P_d(C_{n-1})-P_d(C_{n})\bigg)\left(1+\Omega_n2^{-\frac{B}{N_t-1}}\right)^{-(N_t-1)}=\varepsilon_0$
and then let $B=\lceil B^{\dag}\rceil$ because of its integer
constraint.

\subsection{Noise-Limited Case}
If the noise is quite large, the inter-user interference can be
negligible compared with the noise, namely the noise-limited case.
In this scenario, the received SINR of the $k$th MU can be
approximated as
\begin{equation}
\rho_k\approx\gamma|\textbf{h}_k^H\textbf{w}_k|^2.\label{eqn26}
\end{equation}
Similarly, we have the cdf of $\rho_k$ in this case as
\begin{equation}
F_{\rho_k}(x)=1-\exp\bigg(\frac{x}{\gamma}\bigg).\label{eqn27}
\end{equation}
Hence, we could formulate the joint resource allocation as the
following optimization problem
\begin{eqnarray}
J_3: \min\limits_{P,B} &&\eta=\varphi P+\psi N_tB\label{eqn28}\\
s.t.\quad
&&P_d(C_0)-\sum\limits_{n=1}^{N-1}\bigg(P_d(C_{n-1})-P_d(C_{n})\bigg)\exp\left(-\Omega_nN_t\sigma_n^2/P\right)\leq\varepsilon_0\nonumber\\
&&P\leq P_0\nonumber\\
&&B\leq B_0.\nonumber
\end{eqnarray}

In this case, $B=0$ is the optimal solution to $J_3$, which is
consistent with our intuition. This is because when the noise is
dominant, CSI feedback hardly affects the SINR. Furthermore, the
optimal $P$ can be obtained by solving the function
$P_d(C_0)-\sum\limits_{n=1}^{N-1}\bigg(P_d(C_{n-1})-P_d(C_{n})\bigg)\exp\left(-\Omega_nN_t\sigma_n^2/P\right)=\varepsilon_0$.

\section{Performance Analysis and Simulation Results}
To examine the effectiveness of the proposed cost minimizing joint
transmit power and feedback bandwidth allocation scheme with delay
guarantee, we present several numerical results in different
scenarios. For all scenarios, we set $N_t=4$, $N=8$, $W=1$MHz,
$\sigma_n^2=1$, $P_{\textrm{obj}}=10^{-4}$ and $N_b=1080$ for
convenience. Noticeably, considering resource limitation, we set
$P_0=40$dB and $B_0=10$. The codebooks are designed based on vector
quantization (VQ) method \cite{VQ1} \cite{QCA}.

\begin{figure}[h] \centering
\includegraphics [width=0.8\textwidth] {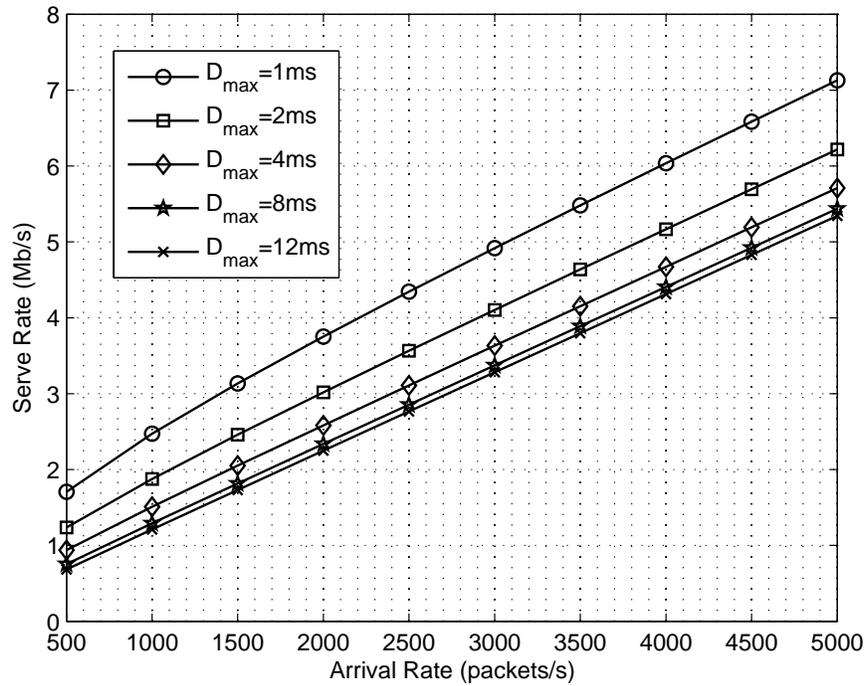}
\caption {Minimum required served rate for high arrival rate.}
\label{Fig2}
\end{figure}

\begin{figure}[h] \centering
\includegraphics [width=0.8\textwidth] {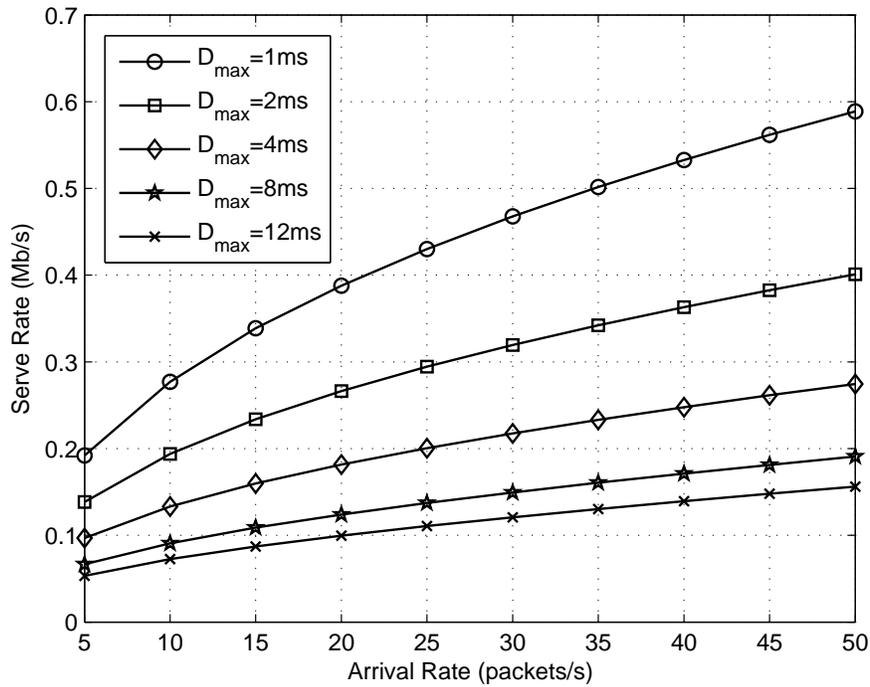}
\caption {Minimum required served rate for low arrival rate.}
\label{Fig3}
\end{figure}

Fig.\ref{Fig2} and Fig.\ref{Fig3} show the required minimum served
rates $C_{\textrm{min}}(\lambda,D_{\textrm{max}})$ for the scenarios
with high and low arrival rates respectively when
$\varepsilon_0=0.01$. Depending on the arrival rates, we can see
that there are different trends for the served rates in these two
scenarios. For high arrival rate $\lambda$, minimum serve rate
increases approximately linearly, which is well consistent with
Theorem 1. However, for low arrival rate, the maximum delay
constraint $D_{\textrm{max}}$ has a great impact on the slope of
variation of the required serve rate, where the slope is inversely
proportional to $D_{\textrm{max}}$. As seen in Fig.\ref{Fig3}, when
$D_{\textrm{max}}$ is greater than $8$ms, the slope is approaching
zero asymptotically. In addition, the gap of the required serve
rates
$\Delta_C=C_{\textrm{min}}(D_{\textrm{max},1})-C_{\textrm{min}}(D_{\textrm{max},2})$
is not a linear function of
$\Delta_D=D_{\textrm{max},1}-D_{\textrm{max},2}$. For example, the
gap of the required serve rates due to the delay relaxation from
$D_{\textrm{max}}=8$ms to $4$ms is larger than the delay relaxation
from $D_{\textrm{max}}=12$ms to $8$ms.

In Tab.\ref{Tab3}, we compare the joint resource allocation results
based on the proposed algorithm and exhaustive search algorithm. For
ease of comparison, we fix arrival rate as $\lambda=300$ packets/s
and the upper bound on average violation probability
$\varepsilon_0=0.01$. In addition, we use $\xi=\psi/\varphi$ to
denote the relative cost factor. For a strict delay constraint, such
as $D_{\textrm{max}}=2$ms, with the increase of the relative cost
factor, the optimal feedback bandwidth decreases accordingly while
the required transmit power increases, this is because as the cost
factor of feedback bandwidth increases, higher transmit power has a
lower total cost while satisfying the delay constraint. Therefore,
we could flexibly adjust the resource combination by changing the
relative cost factor according to the characteristics of the
considered system. Moreover, it is found that the proposed algorithm
gets the same results as exhaustive search algorith, which
reconfirms our theoretical claim. For the case with loose
constraint, e.g. $D_{\textrm{max}}=8$ms, the above observations also
hold true. Comparing with the allocation results with the same
$\xi$, the two cases use the same feedback bandwidth, but the case
with $D_{\textrm{max}}=2$ms has a higher transmit power, since it is
relatively cheaper to add more power than to reduce feedback
bandwidth.

\section{Conclusion}
A major contribution of this paper is to provide a framework of
joint transmit power and feedback bandwidth allocation with delay
guarantee so that the two scarce resources can be utilized in a more
efficient manner according to the characteristics of a multi-user
MIMO broadcast system. First, according to the effective bandwidth
theory, we established the intrinsic relationship between minimum
required serve rate and maximum delay constraint. Then, based on
adaptive modulation, we formulated the average violation probability
with respect to maximum delay in terms of transmit power and
codebook size. Eventually, by minimizing the total resource cost
while satisfying the delay constraint, an optimal joint resource
allocation scheme was derived accordingly.

\begin{appendices}
\section{Proof of Theorem 1}
Given $N_b$, $D_{\textrm{max}}$ and $\varepsilon$, if the arrival
rate is large enough, in order to ensure that $\varepsilon$ is a
positive constant between 0 and 1, $s=\delta(C)$ should be a
positive constant close to zero. Thereby, $\exp(sN_b)$ in
(\ref{eqn9}) can be approximately expressed as $1+N_bs+N_b^2s^2$ by
Taylor expansion at the point zero so that
$C_{\textrm{min}}(D_{\textrm{max}})=\alpha(s)\simeq\lambda
(N_b+N_b^2s/2)$ by substituting $\exp(sN_b)\simeq1+N_bs+N_b^2s^2/2$
into (\ref{eqn9}). Meanwhile, by replacing $C$ in (\ref{eqn7}) with
$\lambda (N_b+N_b^2s/2)$, we have
\begin{equation}
\varepsilon\simeq\hat{\rho}\exp(-\lambda
N_bD_{\textrm{max}}(1+sN_b/2)s).\label{appe1}
\end{equation}
Rearranging (\ref{appe1}), it is obtained that
\begin{equation}
\frac{N_b}{2}s^2+s+\frac{a(\lambda,D_{\textrm{max}})}{N_b}=0,\label{appe2}
\end{equation}
where
$a(\lambda,D_{\textrm{max}})=\frac{\ln\varepsilon-\ln\hat{\rho}}{\lambda
D_{\textrm{max}}}$. By solving the function (\ref{appe2}), we have
\begin{equation}
s=\frac{\sqrt{1-2a(\lambda,D_{\textrm{max}})}-1}{N_b}.\label{appe3}
\end{equation}
Therefore, the required minimum served rate can be approximately
expressed as
\begin{eqnarray}
C_{\textrm{min}}(D_{\textrm{max}})&\simeq&\lambda
(N_b+N_b^2s/2)\nonumber\\
&=&\frac{N_b}{2}\lambda+\frac{N_b}{2}\lambda\sqrt{1-2a(\lambda,D_{\textrm{max}})}\nonumber\\
&\approx&N_b\lambda-\frac{N_b(\ln\varepsilon-\ln\hat{\rho})}{2D_{\textrm{max}}},\label{appe4}
\end{eqnarray}
where (\ref{appe4}) follows from the fact that
$\sqrt{1+x}\approx1+\frac{x}{2}$, if $-1<x\leq1$. Thereby, we
validate the claim of Theorem 1.

\section{Proof of the optimality of joint resource allocation scheme}
Assume $\cup(P_i,B_i)$, $i=0,1,\cdots, B_0$, where $B_i=i$, is the
set of all the feasible solutions of $J_1$. If $B^{\dagger}$ is an
integer, then $(P^{\dagger},B^{\dagger})$ is also the optimal
solution of $J_1$ definitely. Otherwise, we take $B_c=\lceil
B^{\dagger}\rceil$ and $B_f=\lfloor B^{\dagger}\rfloor$ as the two
candidates, and derive the corresponding optimal transmit power
$P_c$ and $P_f$. Since the violation probability is the monotonously
decreasing function of $P$ and $B$, we have $P_c<P_f$. In what
follows, we prove $(P_c, B_c)$ is the optimal solution of $J_1$ if
$\eta(P_c, B_c)<\eta(P_f, B_f)$, otherwise $(P_f, B_f)$ is the
optimal solution. Prior to the proof, we first present the following
lemma:

\emph{Lemma 1}: Given the requirement of the violation probability,
the reduction of transmit power $\triangle P$ becomes smaller
gradually by adding the same feedback bandwidth $\triangle B$ as
feedback bandwidth $B$ increases.

Lemma 1 holds true since the violation probability is a power
function of feedback bandwidth. If $\eta(P_c, B_c)<\eta(P_f, B_f)$,
then the cost of the added 1bit feedback bandwidth is less than that
of the reduced transmit power. Assuming $(P_{f-1},B_{f-1})$ is the
feasible solution of $J_1$, due to $P_{f-1}-P_f>P_f-P_c$ according
to Lemma 1, we have $\eta(P_{f-1},B_{f-1})>\eta(P_f, B_f)$. Thus, it
is obtained that $\eta(P_{i},B_{i})>\eta(P_f, B_f)>\eta(P_c, B_c)$
for all $i<f$. We consider the case of $i>c$. Assuming
$\eta(P_{i},B_{i})<\eta(P_c, B_c)$ and $(P^{'},B^{'})$ is a feasible
solution of the relaxed problem, where
$B^{'}-B^{\dagger}=B_{i}-B_{c}$. According to Lemma 1, we have
$P^{\dagger}-P^{'}>P_{c}-P_{i}$ because of $B^{\dagger}<B_{c}$. If
$\eta(P_{i},B_{i})<\eta(P_c, B_c)$, we have
$\eta(P^{'},B^{'})<\eta(P^{\dagger}, B^{\dagger})$. However,
$(P^{\dagger}, B^{\dagger})$ is the optimal solution of the relaxed
problem, it is impossible to find the $(P^{'},B^{'})$ such that
$\eta(P^{'},B^{'})<\eta(P^{\dagger}, B^{\dagger})$. In other words,
$\eta(P_{i},B_{i})<\eta(P_c, B_c)$ for any $i>c$ can not hold true.
Hence, $(P_c, B_c)$ is the optimal solution of $J_1$. Similarly, if
$\eta(P_c, B_c)>\eta(P_f, B_f)$, $(P_f, B_f)$ is optimal.

\end{appendices}

\begin{table}\centering
\caption{The Important Parameters for Different Modulation Modes}
\label{Tab1}
\begin{tabular}{|c||c|c|c|c|c|c|c|}\hline
 & Mode 1 & Mode 2 & Mode 3 & Mode 4 & Mode 5 & Mode 6 & Mode 7\\
\hline\hline Modulation & BPSK & QPSK & 8QAM & 16QAM & 32QAM & 64QAM
& 128QAM\\\hline Rate (bits/sym.) & 1 & 2 & 3 & 4 & 5 & 6 &
7\\\hline $a_n$ & 67.7328 & 73.8279 & 58.7332 & 55.9137 & 50.0552 &
42.5594 & 40.2559 \\\hline $g_n$ & 0.9819 & 0.4945 & 0.1641 & 0.0989
& 0.0381 & 0.0235 & 0.0094\\\hline $\gamma_{pn} (dB)$ & 6.3281 &
9.3945 & 13.9470 & 16.0938 & 20.1103 & 22.0340 & 25.9677\\\hline

\end{tabular}
\end{table}

\begin{table}\centering
\caption{Resource Allocation Results for Different Relative Cost
Factor} \label{Tab3}
\begin{tabular}{|c|c|c|c|c|c|c|c|}\hline
& & $\xi$ & 80 & 120 & 160 & 200 & 240 \\
\hline\hline & & $B$ & 10 & 8 & 7 & 6 & 5
\\\cline{3-8} $D_{\textrm{max}}=2$ms & $\textmd{Exhaustive Search Algorithm}$& $P$(dB) & 38.0 & 38.6 & 38.9 & 39.3 & 39.8\\
\cline{2-8} & & $B$ & 10 & 8 & 7 & 6 & 5
\\\cline{3-8} &$\textmd{Proposed Algorithm}$& $P$(dB) & 38.0 & 38.6 & 38.9 & 39.3 & 39.8\\
\cline{3-8} \hline\hline & & $B$ & 10 & 8 & 7 & 6 & 5
\\\cline{3-8} $D_{\textrm{max}}=8$ms & $\textmd{Exhaustive Search Algorithm}$& $P$(dB) & 37.7 & 38.2 & 38.6 & 39.0 & 39.4\\
\cline{2-8} & & $B$ & 10 & 8 & 7 & 6 & 5
\\\cline{3-8} &$\textmd{Proposed Algorithm}$& $P$(dB) & 37.7 & 38.2 & 38.6 & 39.0 & 39.4\\
\cline{3-8} \hline
\end{tabular}
\end{table}

\end{spacing}
\end{document}